\documentclass[aps,twocolumn,showpacs,preprintnumbers,amsmath,amssymb,floatfix]{revtex4}
\usepackage{graphicx}
\usepackage{dcolumn}
\usepackage{bm}
\usepackage{longtable}

\begin{document}
\title{Geometric Random Inner Products: A New Family of Tests for Random
Number Generators}
\author{Shu-Ju Tu}
\email{sjtu@physics.purdue.edu}
\affiliation{Department of Physics, Purdue University, West Lafayette, Indiana
47907-1396}
\author{Ephraim Fischbach}
\email{ephraim@physics.purdue.edu}
\affiliation{Department of Physics, Purdue University, West Lafayette, Indiana
47907-1396}
\date{\today}

\begin{abstract}
We present a new computational scheme, GRIP (Geometric Random Inner
Products), for testing the quality of random number generators. The
GRIP formalism utilizes geometric probability techniques to calculate
the average scalar products of random vectors generated in geometric
objects, such as circles and spheres. We show that these average scalar
products define a family of geometric constants which can be used
to evaluate the quality of random number generators. We explicitly
apply the GRIP tests to several random number generators frequently
used in Monte Carlo simulations, and demonstrate a new statistical
property for good random number generators.
\end{abstract}
\pacs{02.50.Ng}

\maketitle

\section{Introduction}

Monte Carlo methods are among the most widely used numerical algorithms
in computational science and engineering~\cite{top10}. The key element
in a Monte Carlo calculation is the generation of random numbers.
Although a truly random number sequence produced by either a physical
process such as nuclear decay, an electronic device etc., or by a
computer algorithm, may not actually exist, a new and computationally
easy-to-implement scheme to investigate random number generators is
always highly desirable.

There have been many proposed schemes for the quality
measure of random number generators~\cite{Knuth,recipes,Marsaglia_1,garcia,Giordano,Gentle,ferrenberg,vattulaninen}.
These computational tests are based either on probability theory and
statistical methods (for example: the \( \chi ^{2} \) test, the Smirnov-Kolmogorov
test, the correlation test, the spectral test, and the DieHard battery
of randomness tests), or on mathematical modeling and simulation for
physical systems (for example: random walks and Ising model simulations).
These methods also open the door to studying the properties of random
number sequences such as randomness and complexity~\cite{rand}. Some
important attempts at an operational definition of randomness were
previously developed by Kolmogorov and Chaitin (algorithmic informational
theory)~\cite{kolmogorov,chaitin,chaitin_001,chaitin_002} and by
Pincus (approximate entropy)~\cite{pincus}.

In this paper, we study a new method to measure \( n \)-dimensional
randomness which we denote by GRIP (Geometric Random Inner Products).
The GRIP family of tests is based on the observation that the average
scalar products of random vectors produced in geometric objects (e.g.,
circles and spheres), define geometric constants which can be used
to evaluate the quality of random number generators. After presenting
the simplest example of a GRIP test, we exhibit a computational
method for implementing GRIP, which is then used to analyze a number
of random number generators. We then discuss the GRIP formalism in
detail and show how a random number sequence, when converted to random
points in a space defined by a geometric object, can produce a series
of known geometric constants. Later we introduce additional members
and include them within the GRIP family. We then present the computational
results for configurations of four, six, and eight random points,
along with a consideration of some key issues. Finally, we conclude by discussing
how the GRIP test measures the quality of random number generators
by explicitly adding a new quantitative property to random number
sequences along with the three known qualitative properties summarized
in Ref.~\cite{rand}.

\section{General Description of the GRIP Formalism}

The GRIP scheme is derived from the theory of random distance distribution
for spherical objects, and can be generalized to other geometric objects
with arbitrary densities~\cite{sjtu0,sjtu1}. First, three random
points (\( \vec{r}_{1} \), \( \vec{r}_{2} \), and \( \vec{r}_{3} \))
are independently produced from the sample space defined by a geometric
object. We then evaluate the average inner product of \( \vec{r}_{12}\cdot \vec{r}_{23} \)
from two associated random vectors, \( \vec{r}_{12}=\vec{r}_{2}-\vec{r}_{1} \)
and \( \vec{r}_{23}=\vec{r}_{3}-\vec{r}_{2} \). For a geometric object
such as an \( n \)-ball of uniform density with a radius \( R \),
the analytical result is a geometric constant which can be expressed
in terms of the dimensionality \( n \) of the space~\cite{sjtu0,sjtu1}:\begin{equation}
\label{eq_master}
\left\langle \vec{r}_{12}\cdot \vec{r}_{23}\right\rangle _{n}=-\frac{n}{n+2}R^{2}.
\end{equation}
 A simple derivation of Eq.~(\ref{eq_master}) can be found in the
Appendix.

The following procedures are the numerical implementation of our testing
programs. A random number sequence produced from a random number generator
is used to generate a series of three random points \( \vec{r}_{1} \),
\( \vec{r}_{2} \), and \( \vec{r}_{3} \) such that these random
points are uniformly distributed in an \( n \)-dimensional spherical
ball \( \mathbf{B} \) of radius \( R \), where \begin{equation}
\mathbf{B}=\left\{ \left( x_{1},x_{2},\cdots ,x_{n}\right) :x_{1}^{2}+x_{2}^{2}+\cdots +x_{n}^{2}\leq R^{2}\right\} .
\end{equation}
 We then compute a series of values for \( \vec{r}_{12}\cdot \vec{r}_{23} \).
If \( \vec{r}_{12}\cdot \vec{r}_{23} \) is evaluated \( N \) times,
then statistically we expect \begin{equation}
\label{eq_n_001}
\lim _{N\rightarrow \infty }\frac{1}{N}\sum _{i=1}^{N}\left( \vec{r}_{12}\cdot \vec{r}_{23}\right) _{i}=-\frac{n}{n+2}R^{2},
\end{equation}
 as predicted by Eq.~(\ref{eq_master}).

\section{Random Number Generators}

We now apply the GRIP test to the following random number generators
frequently used in Monte Carlo simulations. 

\begin{enumerate}
\item RAN0 - a linear congruential generator~\cite{Knuth,recipes}: \begin{equation}
x_{n}=16807\times x_{n-1}\quad \mathrm{mod}\quad 2147483647.
\end{equation}

\item RAN3 - a lagged Fibonacci generator~\cite{Knuth,recipes}: \begin{equation}
x_{n}=\left( x_{n-55}-x_{n-24}\right) \quad \mathrm{mod}\quad 2^{31}.
\end{equation}

\item R31 - a generalized feedback shift register (GFSR) generator~\cite{Knuth,ferrenberg,vattulaninen}:
\begin{equation}
x_{n}=x_{n-31}\oplus x_{n-3},
\end{equation}
 where \( \oplus  \) is the bitwise exclusive OR operator. 
\item durxor - a generator selected from IBM ESSL (Engineering and Scientific
Subroutine Library)~\cite{essl}. 
\item durand - a generator selected from IBM ESSL (Engineering and Scientific
Subroutine Library) and the sequence period of durand is shorter than
durxor~\cite{essl}. 
\item ran\_gen - one of the subroutines in IMSL libraries from Visual Numeric~\cite{imsl}. 
\item Random - a Fortran 90/95 standard intrinsic random number generator~\cite{metcalf}. 
\item Weyl - a Weyl sequence generator~\cite{holian,tretiakov}: \begin{equation}
x_{n}=\left\{ n\alpha \right\} ,
\end{equation}
 where \( \left\{ x\right\}  \) is the fractional part of \( x \),
and \( \alpha  \) is an irrational number such as \( \sqrt{2} \). 
\item NWS - a nested Weyl sequence generator~\cite{holian,tretiakov}: \begin{equation}
x_{n}=\left\{ n\left\{ n\alpha \right\} \right\} .
\end{equation}

\item SNWS - a shuffled nested Weyl sequence generator~\cite{holian,tretiakov}:
\begin{eqnarray}
s_{n} & = & M\left\{ n\left\{ n\alpha \right\} \right\} +\frac{1}{2},\\
x_{n} & = & \left\{ s_{n}\left\{ s_{n}\alpha \right\} \right\} ,
\end{eqnarray}
 where \( M \) is a large positive integer. 
\end{enumerate}
The computational results obtained from Eq.~(\ref{eq_n_001}) when \( n=3 \)
and \( n=9 \) are presented in Table~\ref{table_3_001}. Results
for random number generators based on other algorithms such as the
Data Encryption Standard (DES)~\cite{Knuth,recipes} can be found
in Ref.~\cite{sjtu2} along with the computed results obtained from
other geometric objects. We note that both the ran\_gen and RAN0 generators perform better overall,
while the NWS and Weyl generators (which are based on the Weyl sequence method) are ranked lowest 
compared to the other generators.
The reasons why this is the case will be discussed later.

\begin{table*}
\caption{Computed results for \protect\( \left\langle \vec{r}_{12}\cdot \vec{r}_{23}\right\rangle _{n}\protect \).
RNG denotes the specific random number generator defined in the text,
and ``Expected'' is the exact result obtained from Eq.~(\ref{eq_master}).
The entries are ranked in terms of their errors, which are the absolute
values of the differences between the expected and computed results.
For each entry in the table \protect\( N=10^{8}\protect \), and multiple seeds were used where appropriate.
\label{table_3_001} }

\begin{ruledtabular}
\begin{tabular}{ccccccc}
Rank&RNG&\( n=3 \)&Error&RNG&\( n=9 \)&Error\\
\hline
1&ran\_gen&\( -0.59999802 \)&\( 0.00000198 \)&RAN0&\( -0.81819136 \)&\( 0.00000955 \)\\
2&RAN0&\( -0.60000722 \)&\( 0.00000722 \)    &ran\_gen&\( -0.81821041 \)&\( 0.00002860 \)\\
3&R31&\( -0.60005031 \)&\( 0.00005031 \)     &Random&\( -0.81821550 \)&\( 0.00003369 \)\\
4&durand&\( -0.59991945 \)&\( 0.00008055 \)  &durand&\( -0.81821772 \)&\( 0.00003591 \)\\
5&durxor&\( -0.59991306 \)&\( 0.00008694 \)  &durxor&\( -0.81822185 \)&\( 0.00004004 \)\\
6&RAN3&\( -0.59988610 \)&\( 0.00011390 \)    &R31&\( -0.81824459 \)&\( 0.00006278 \)\\ 
7&Random&\( -0.59987912 \)&\( 0.00012088 \)  &RAN3&\( -0.81827541 \)&\( 0.00009360 \)\\
8&SNWS&\( -0.59969277 \)&\( 0.00030723 \)    &SNWS&\( -0.81795246 \)&\( 0.00022935 \)\\ 
9&NWS&\( -0.62988317 \)&\( 0.02988317 \)     &NWS&\( -0.82539808 \)&\( 0.00721627 \)\\
10&Weyl&\( -1.80809907 \)&\( 1.20809907 \)   &Weyl&\( -1.05604651 \)&\( 0.23786470 \)\\ 
\hline
&{\bf Expected}&\( -0.60000000 \)&&{\bf Expected}&\( -0.81818181 \)&\\
\end{tabular}
\end{ruledtabular}
\end{table*}

\section{GRIP Analysis}

In the following, we analyze the relationship between GRIP and a random
number sequence, and show how a good random number sequence, when
converted to random points in a a space defined by a geometric object,
can produce a series of known \( n \)-dimensional geometric constants.
A random number sequence generated from a random number generator
can be written as, \begin{equation}
\label{eq_sequence}
a_{1}a_{2}a_{3}a_{4}a_{5}a_{6}a_{7}a_{8}a_{9}a_{10}\cdots \cdots .
\end{equation}
 When the sequence is converted to represent random points in a \( 2 \)-dimensional
geometric object, the random numbers in Eq.~(\ref{eq_sequence}) can
then be grouped in pairs as \begin{equation}
\left( a_{1}a_{2}\right) \left( a_{3}a_{4}\right) \left( a_{5}a_{6}\right) a_{7}a_{8}a_{9}a_{10}\cdots \cdots ,
\end{equation}
 where Cartesian coordinates are used. The first set of random points
\( \left\{ \vec{r}_{1},\vec{r}_{2},\vec{r}_{3}\right\}  \) can thus
be identified as \begin{equation}
\vec{r}_{1}=\left( a_{1},a_{2}\right) ,\quad \vec{r}_{2}=\left( a_{3},a_{4}\right) ,\quad \vec{r}_{3}=\left( a_{5},a_{6}\right) .
\end{equation}
 GRIP then uses \( \vec{r}_{1} \), \( \vec{r}_{2} \), and \( \vec{r}_{3} \)
to evaluate the average scalar product which can be computed by rewriting,
\begin{widetext}
\begin{equation}
\label{eq_circle}
\left\langle \vec{r}_{12}\cdot \vec{r}_{23}\right\rangle =\frac{1}{N}\sum ^{N}_{i=1}\sum _{j=1}^{2}\left( a_{6i-4+j}-a_{6i-6+j}\right) \left( a_{6i-2+j}-a_{6i-4+j}\right) ,
\end{equation}
\end{widetext}
 where \( N \) is a large positive integer. When the geometric object
is a circle of radius \( R \) and uniform density, we expect \( \left\langle \vec{r}_{12}\cdot \vec{r}_{23}\right\rangle \approx -0.5R^{2} \)
as predicted by Eq.~(\ref{eq_master}).

The analysis for \( 2 \)-dimensional GRIP can be immediately generalized
to the \( n \)-dimensional case. When the sequence in Eq.~(\ref{eq_sequence})
is used to generate random points in a \( n \)-dimensional spherical
object, we can regroup Eq.~(\ref{eq_sequence}) as follows: 
\begin{widetext}
\begin{equation}
\left( a_{1}\cdots a_{k}\right) \left( a_{k+1}\cdots a_{2k}\right) \left( a_{2k+1}\cdots a_{3k}\right) \left( \cdots \right) \left( \cdots \right) \left( \cdots \right) \cdots \cdots .
\end{equation}
\end{widetext}
 The average scalar product of \( \vec{r}_{12}\cdot \vec{r}_{23} \)
can then be expressed as 
\begin{widetext}
\begin{equation}
\label{eq_500}
\left\langle \vec{r}_{12}\cdot \vec{r}_{23}\right\rangle =\frac{1}{N}\sum ^{N}_{i=1}\sum _{j=1}^{n}\left( a_{3in-2n+j}-a_{3in-3n+j}\right) \left( a_{3in-n+j}-a_{3in-2n+j}\right) .
\end{equation}
\end{widetext}
 When the geometric object is an \( n \)-ball with a radius \( R=1 \)
and a uniform density, we expect from Eq.~(\ref{eq_master}) that
the result of Eq.~(\ref{eq_500}) should be a geometric constant,
\( -n/(n+2) \).

\section{GRIP Members}

For practical computational purposes, we may wish to transform a random
number sequence from a uniform density distribution to one which is
non-uniform. One of the most important non-uniform density distributions
is the Gaussian (normal) distribution \( P(r) \) with mean zero and
standard deviation \( \sigma  \), \begin{equation}
\label{eq_Gaussian}
P(r)=\frac{1}{(2\pi )^{n/2}\sigma ^{n}}e^{-(1/2)\left( r^{2}/\sigma ^{2}\right) }.
\end{equation}
 Here \( \int _{0}^{\infty }P(r)\, dr=1 \), \( r=\left( x_{1}^{2}+\cdots +x_{n}^{2}\right) ^{1/2} \),
and \( n \) is the space dimensionality. One can use either the Box-Muller
transformation method to generate a random number sequence with a
Gaussian density distribution, or use available subroutines from major
computational scientific libraries such as IBM ESSL and IMSL~\cite{essl,imsl}.
By applying the probability density function of the random distance
distribution as discussed in Ref.~\cite{sjtu1}, one can add a new
GRIP member to investigate the quality of a Gaussian random number
generator, and this new GRIP test can be expressed as: \begin{equation}
\label{eq_a_001}
\left\langle \vec{r}_{12}\cdot \vec{r}_{23}\right\rangle _{n}=-n\sigma ^{2}.
\end{equation}
 A very common situation arises when one has to produce random points
uniformly distributed on the surface of an \( n \)-sphere of radius
\( R \). Some general computational techniques for doing this are
summarized in Refs.~\cite{Knuth,sjtu0}. We can then use \begin{equation}
\label{eq_a_002}
\left\langle \vec{r}_{12}\cdot \vec{r}_{23}\right\rangle _{n}=-R^{2},
\end{equation}
 to examine the quality of such transformed random number generators
as discussed in Ref.~\cite{sjtu2}.

Another application of the GRIP formalism is in stochastic geometry.
We can design a test scheme for a configuration utilizing any number
of random points~\cite{sjtu2}, and these tests can be included in
the GRIP family. Among the tests are: 

\begin{enumerate}
\item Four uniform random points configuration for an \( n \)-ball of radius
\( R \) 
\begin{eqnarray}
\left\langle \left( \vec{r}_{12}\cdot \vec{r}_{23}\right) \left( \vec{r}_{34}\cdot \vec{r}_{41}\right) \right\rangle _{n} & = & \frac{n\left( n+1\right) }{\left( n+2\right) ^{2}}R^{4},\label{eq_b_001} \\
\left\langle \left( \vec{r}_{12}\cdot \vec{r}_{34}\right) \left( \vec{r}_{23}\cdot \vec{r}_{41}\right) \right\rangle _{n} & = & \frac{2n}{\left( n+2\right) ^{2}}R^{4},\label{eq_b_002} \\
\left\langle \vec{r}_{13}\cdot \vec{r}_{24}\right\rangle _{n} & = & 0.\label{eq_b_003} 
\end{eqnarray}

\item \( 2m \) uniform random points configuration for an \( n \)-ball
of radius \( R \) 
\begin{widetext}
\begin{equation}
\label{eq_b_004}
\left\langle \left( \vec{r}_{12}\cdot \vec{r}_{23}\right) \cdots \cdots \left( \vec{r}_{2m-1\, \, 2m}\cdot \vec{r}_{2m\, \, 1}\right) \right\rangle _{n}=(-1)^{m}\frac{n\left( n^{m-1}+1\right) }{\left( n+2\right) ^{m}}R^{2m},
\end{equation}
\end{widetext}
 where \( 2m \) (\( m=2, \) \( 3 \), \( 4 \) etc.) is a positive
even number. 
\end{enumerate}
A derivation of Eq.~(\ref{eq_b_001}) can be found in the Appendix.
We summarize the computational results for Eq.~(\ref{eq_b_004}) when
\( m=2, \) \( 3, \) \( 4 \) in in Tables~\ref{table_4_001}, \ref{table_6_001},
and \ref{table_8_001}. A discussion of other results, such as Eqs.~(\ref{eq_a_001})
and (\ref{eq_a_002}), can be found in Ref.~\cite{sjtu2}.

We observe that all of the generators except NWS and Weyl perform significantly better 
in $n=3$ than in $n=9$
using the GRIP test based on   
$\left\langle \left( \vec{r}_{12}\cdot \vec{r}_{23}\right) \left( \vec{r}_{34}\cdot 
\vec{r}_{41}\right) \right\rangle_{n}$.
We also note from Table~\ref{table_4_001}, 
and the $n=9$ results (from R31 to RAN0), that these results are clearly biased to larger numbers 
compared to the expected value. One interpretation may be that 
$\left\langle \left( \vec{r}_{12}\cdot \vec{r}_{23}\right) \left( \vec{r}_{34}\cdot 
\vec{r}_{41}\right) \right\rangle_{9}$ is a more sensitive and dedicated computational test
for investigating random number generators than other GRIP tests.
We also note that the results for $n=9$ are overall worse than $n=3$, and that the results for
$\left\langle \left( \vec{r}_{12}\cdot \vec{r}_{23}\right) \left( \vec{r}_{34}\cdot 
\vec{r}_{41}\right) \right\rangle_{9}$ reveal a more significant bias than in any of the other cases.
These results suggest that the GRIP test either in higher dimensions (large $n$), or using a 
configuration of 
four random points, can serve as a more computationally sensitive test to detect non-random patterns 
hidden in random number sequences.
Finally we note that it is not surprising that the NWS and Weyl generators are ranked worst 
among all cases in our GRIP test. 
As reported previously in~\cite{tretiakov}, these two show unacceptable
non-random behavior and strong correlations. 

\begin{table*}
\caption{Computed results for \protect\( \left\langle \left( \vec{r}_{12}\cdot \vec{r}_{23}\right) \left( \vec{r}_{34}\cdot \vec{r}_{41}\right) \right\rangle _{n}\protect \).
RNG denotes the specific random number generator defined in the text,
and ``Expected'' is the exact result obtained from Eq.~(\ref{eq_master}).
The entries are ranked in terms of their errors, which are the absolute
values of the differences between the expected and computed results.
For each entry in the table \protect\( N=10^{8}\protect \), and multiple seeds were used where appropriate.
\label{table_4_001}}
\begin{ruledtabular}
\begin{tabular}{ccccccc}
Rank&RNG&\( n=3 \)&Error&RNG&\( n=9 \)&Error\\
\hline
1&RAN3&\( 0.47999737 \)&\( 0.00000263 \)&R31&\( 0.74538039 \)&\( 0.00157874 \)\\
2&RAN0&\( 0.47995440 \)&\( 0.00004560 \)&RAN3&\( 0.74567962 \)&\( 0.00187797 \)\\
3&durxor&\( 0.48004715 \)&\( 0.00004715 \)&SNWS&\( 0.74582752 \)&\( 0.00202587 \)\\
4&durand&\( 0.48006530 \)&\( 0.00006530 \)&ran\_gen&\( 0.74634598 \)&\( 0.00254433 \)\\
5&R31&\( 0.48008265 \)&\( 0.00008265 \)&durxor&\( 0.74637990 \)&\( 0.00257825 \)\\
6&Random&\( 0.47990347 \)&\( 0.00009653 \)&Random&\( 0.74644366 \)&\( 0.00264201 \)\\
7&ran\_gen&\( 0.47986482 \)&\( 0.00013518 \)&durand&\( 0.74646979 \)&\( 0.00266814 \)\\
8&SNWS&\( 0.47975570 \)&\( 0.00024430 \)&RAN0&\( 0.74659547 \)&\( 0.00279382 \)\\
9&NWS&\( 0.55841828 \)&\( 0.07842818 \)&NWS&\( 0.69652947 \)&\( 0.04727218 \)\\
10&Weyl&\( 3.31162983 \)&\( 2.83162983 \)&Weyl&\( 1.28608478 \)&\( 0.54228305 \)\\
\hline 
&{\bf Expected}&\( 0.48000000 \)&&{\bf Expected}&\( 0.74380165 \)&\\
\end{tabular}
\end{ruledtabular}
\end{table*}

\begin{table*}
\caption{Computed results for \protect\( \left\langle \left( \vec{r}_{12}\cdot \vec{r}_{23}\right) \left( \vec{r}_{34}\cdot \vec{r}_{45}\right) \left( \vec{r}_{56}\cdot \vec{r}_{61}\right) \right\rangle _{n}\protect \).
RNG denotes the specific random number generator defined in the text,
and ``Expected'' is the exact result obtained from Eq.~(\ref{eq_master}).
The entries are ranked in terms of their errors, which are the absolute
values of the differences between the expected and computed results.
For each entry in the table \protect\( N=10^{8}\protect \), and multiple seeds were used where appropriate.
\label{table_6_001}}
\begin{ruledtabular}
\begin{tabular}{ccccccc}
Rank&RNG&\( n=3 \)&Error&RNG&\( n=9 \)&Error\\
\hline
1&durand&\( -0.24000387 \)&\( 0.00000387 \)&SNWS&\( -0.55453884 \)&\( 0.00006852 \)\\ 
2&Random&\( -0.24001846 \)&\( 0.00001846 \)&R31&\( -0.55373136 \)&\( 0.00073896 \)\\
3&RAN3&\( -0.23997958 \)&\( 0.00002042 \)&ran\_gen&\( -0.55521687 \)&\( 0.00074655 \)\\
4&ran\_gen&\( -0.24003206 \)&\( 0.00003206 \)&durxor&\( -0.55524213 \)&\( 0.00077181 \)\\
5&RAN0&\( -0.23994286 \)&\( 0.00005714 \)&durand&\( -0.55528032 \)&\( 0.00081000 \)\\
6&durxor&\( -0.24011639 \)&\( 0.00011639 \)&Random&\( -0.55533746 \)&\( 0.00086714 \)\\
7&SNWS&\( -0.23964945 \)&\( 0.00035055 \)&RAN0&\( -0.55561838 \)&\( 0.00114806 \)\\
8&R31&\( -0.24070892 \)&\( 0.00070892 \)&RAN3&\( -0.55595656 \)&\( 0.00148624 \)\\
9&NWS&\( -0.28813072 \)&\( 0.04813072 \)&NWS&\( -0.48246750 \)&\( 0.07200282 \)\\
10&Weyl&\( -5.78662461 \)&\( 5.54662461 \)&Weyl&\( -1.31133451 \)&\( 0.75686419 \)\\
\hline
&{\bf Expected}&\( -0.24000000 \)&&{\bf Expected}&\( -0.55447032 \)&\\
\end{tabular}
\end{ruledtabular}
\end{table*}

\begin{table*}
\caption{Computed results for \protect\( \left\langle \left( \vec{r}_{12}\cdot \vec{r}_{23}\right) \left( \vec{r}_{34}\cdot \vec{r}_{45}\right) \left( \vec{r}_{56}\cdot \vec{r}_{67}\right) \left( \vec{r}_{78}\cdot \vec{r}_{81}\right) \right\rangle _{n}\protect \).
RNG denotes the specific random number generator defined in the text,
and {}``Expected'' is the exact result obtained from Eq.~(\ref{eq_master}).
The entries are ranked in terms of their errors, which are the absolute
values of the differences between the expected and computed results.
For each entry in the table \protect\( N=10^{8}\protect \), and multiple seeds were used where appropriate.
\label{table_8_001}}
\begin{ruledtabular}
\begin{tabular}{ccccccc}
Rank&RNG&\( n=3 \)&Error&RNG&\( n=9 \)&Error\\
\hline 
1&ran\_gen&\( 0.13440377 \)&\( 0.00000377 \)&durxor&\( 0.44877762 \)&\( 0.00003778 \)\\
2&durxor&\( 0.13439078 \)&\( 0.00000922 \)&ran\_gen&\( 0.44887037 \)&\( 0.00013053 \)\\
3&RAN3&\( 0.13441796 \)&\( 0.00001796 \)&durand&\( 0.44889786 \)&\( 0.00015802 \)\\
4&RAN0&\( 0.13442104 \)&\( 0.00002104 \)&Random&\( 0.44896356 \)&\( 0.00022372 \)\\
5&durand&\( 0.13437131 \)&\( 0.00002869 \)&RAN3&\( 0.44910757 \)&\( 0.00036773 \)\\
6&Random&\( 0.13430897 \)&\( 0.00009103 \)&RAN0&\( 0.44924443 \)&\( 0.00050459 \)\\
7&SNWS&\( 0.13415229 \)&\( 0.00024771 \)&SNWS&\( 0.44789799 \)&\( 0.00084185 \)\\
8&R31&\( 0.13684585 \)&\( 0.00244585 \)&R31&\( 0.44684269 \)&\( 0.00189715 \)\\
9&NWS&\( 0.16328766 \)&\( 0.02888766 \)&NWS&\( 0.46587567 \)&\( 0.01713583 \)\\
10&Weyl&\( 10.1762479 \)&\( 10.0418479 \)&Weyl&\( 1.53996230 \)&\( 1.09122246 \)\\
\hline
&{\bf Expected}&\( 0.13440000 \)&&{\bf Expected}&\( 0.44873984 \)&\\
\end{tabular}
\end{ruledtabular}
\end{table*}

\section{Conclusions }

We have presented a new computational paradigm for evaluating
the quality of random number generators. We
demonstrate how GRIP helps to understand complexity and randomness
by adding a new property, besides three known properties (typical,
chaotic, and the stability of frequencies)~\cite{rand}, for random
number sequences. This quantitative feature shows how a random number
sequence, when converted to random points in a space defined by a
geometric object, can produce a series of known geometric constants.
Ten random number generators were selected to run our GRIP tests,
and they are ranked based on the errors between the numerical and
analytical results. Finally we note that one implication of our work is that 
computational scientists
should test the random number generators they use in
their simulations, and verify that their random number generators
pass as many proposed tests as possible. 

\appendix*

\section{Derivation of \( \left\langle \left( \vec{r}_{12}\cdot \vec{r}_{23}\right) \right\rangle _{n} \) 
and
\( \left\langle \left( \vec{r}_{12}\cdot \vec{r}_{23}\right) \left( \vec{r}_{34}\cdot \vec{r}_{41}\right) \right\rangle _{n} \)}~\label{derivation_001}

\begin{figure}
{\centering \resizebox*{!}{2in}{\includegraphics{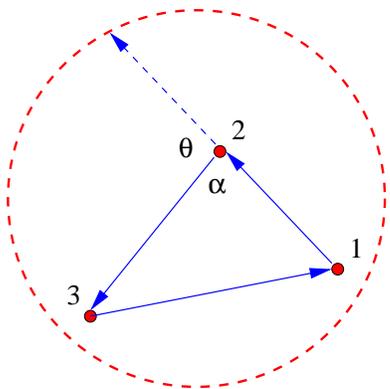}} \par}
\caption{Three random points configuration in a circle.\label{fig_proof}}
\end{figure}

We derive the analytical result of Eq.~(\ref{eq_master}) for a circle
(\( n=2 \)) of radius \( R \) and uniform density. The same derivation
can be applied to the case of \( n \) dimensions where \( n\geq 3 \).
We label three independent random points as \( 1 \), \( 2 \), and
\( 3 \) in Fig.~\ref{fig_proof}, and then calculate \begin{equation}
\label{eq_app_001}
\vec{r}_{12}\cdot \vec{r}_{23}=r_{12}r_{23}\cos \theta =-r_{12}r_{23}\cos \alpha ,
\end{equation}
 where \( \alpha +\theta =\pi  \). From the triangle formed by
the random points, we then have \begin{equation}
\label{eq_app_002}
r_{31}^{2}=r_{12}^{2}+r_{23}^{2}-2r_{12}r_{23}\cos \alpha .
\end{equation}
 Extending this \( 2- \)dimensional case to the \( n \)-dimensional
case, and combining Eqs~(\ref{eq_app_001}) and (\ref{eq_app_002}),
we then evaluate 
\begin{widetext}
\begin{equation}
\label{eq_app_003}
\left\langle \vec{r}_{12}\cdot \vec{r}_{23}\right\rangle _{n}=-\frac{1}{2}\left\langle r_{12}^{2}+r_{23}^{2}-r_{31}^{2}\right\rangle _{n}=-\frac{1}{2}\left\langle r^{2}\right\rangle _{n}=-\frac{1}{2}\int _{0}^{2R}P_{n}(r)r^{2}\, dr=-\frac{n}{n+2}R^{2},
\end{equation}
\end{widetext}
 where we have utilized the fact that \( \vec{r}_{12} \), \( \vec{r}_{23} \),
and \( \vec{r}_{31} \) are three independent random vectors. The
functions \( P_{n}(r) \) in Eq.~(\ref{eq_app_003}), which can be
found in Refs.~\cite{sjtu0,sjtu1,Kendall,Santalo,Solomon,Ambartzumian,Klain,ephraim},
are the probability density functions for the random distance \( r \)
between two random points in an \( n \)-dimensional spherical ball
of radius \( R \) and uniform density.

We consider next the analytical result in Eq.~(\ref{eq_b_001}) for
a circle (\( n=2 \)) of radius \( R \) and uniform density. A similar
derivation can lead to Eqs.~(\ref{eq_b_002}), (\ref{eq_b_003}),
and (\ref{eq_b_004}), as well as to the case of \( n \) dimensions
where \( n\geq 3 \). We begin by expressing \( 4 \) random points
\( \vec{r}_{1} \), \( \vec{r}_{2} \), \( \vec{r}_{3} \), and \( \vec{r}_{4} \)
in Cartesian coordinates, where \( \vec{r}_{i}=\left( x_{i},y_{i}\right)  \).
The expression in Eq.~(\ref{eq_b_001}) can then be evaluated by writing
\begin{widetext}
\begin{equation}
\left\langle \left( \vec{r}_{12}\cdot \vec{r}_{23}\right) \left( \vec{r}_{34}\cdot \vec{r}_{41}\right) \right\rangle _{2}=\frac{\int _{-R}^{R}\, dx_{1}\int _{-\sqrt{R^{2}-x_{1}^{2}}}^{\sqrt{R^{2}-x_{1}^{2}}}\, dy_{1}\cdots \int _{-R}^{R}\, dx_{4}\int _{-\sqrt{R^{2}-x_{4}^{2}}}^{\sqrt{R^{2}-x_{4}^{2}}}\, f_{1}\times f_{2}\, dy_{4}}{\int _{-R}^{R}\, dx_{1}\int _{-\sqrt{R^{2}-x_{1}^{2}}}^{\sqrt{R^{2}-x_{1}^{2}}}\, dy_{1}\cdots \int _{-R}^{R}\, dx_{4}\int _{-\sqrt{R^{2}-x_{4}^{2}}}^{\sqrt{R^{2}-x_{4}^{2}}}\, dy_{4}}=\frac{3}{8}R^{4},
\end{equation}
\end{widetext}
 where \begin{eqnarray*}
f_{1} & = & \left( x_{2}-x_{1}\right) \left( x_{3}-x_{2}\right) +\left( y_{2}-y_{1}\right) \left( y_{3}-y_{2}\right) ,\\
f_{2} & = & \left( x_{4}-x_{3}\right) \left( x_{1}-x_{4}\right) +\left( y_{4}-y_{3}\right) \left( y_{1}-y_{4}\right) .
\end{eqnarray*}
 A derivation of the general result using the probability density functions
\( P_{n}(r) \) in Eq.~(\ref{eq_app_003}) can be found in Ref.~(\cite{sjtu2}).

\begin{acknowledgments}
The authors wish to thank T. K. Kuo and Dave Seaman for helpful discussions
and the Purdue University Computing Center for computing support.
This work was supported in part by the US Department of Energy contract
DE-AC02-76ER1428.
\end{acknowledgments}

\bibliography{references_GRIP}

\end{document}